\documentclass[sigconf,]{acmart}

\setcopyright{none}
\renewcommand\footnotetextcopyrightpermission[1]{}

\usepackage{algorithm}

\usepackage{amsmath,amssymb,amsfonts}
\usepackage{algorithmic}
\usepackage{graphicx}
\usepackage{textcomp}
\usepackage{float}
\usepackage{subfigure}
\usepackage{url}
\usepackage{framed}
\usepackage{multirow}
\usepackage{tikz}
\usepackage{enumitem}
\usepackage{hyperref}
\usepackage{amsthm}
\usepackage{url}

\newcounter{ctr}
\newcommand{\stp}{\addtocounter{ctr}{1}\arabic{ctr}.}
\begin{document}

\title{Fuzzy Logic Guided Reward Function Variation: An Oracle for Testing Reinforcement Learning Programs}

\author{Shiyu Zhang}
\email{shiyu-comp.zhang@connect.polyu.hk}
\affiliation{%
  \institution{The Hong Kong Polytechnic University}
  \city{Hong Kong}
  \country{China}
}

\author{Haoyang Song}
\email{haoyang.song@connect.polyu.hk}
\affiliation{%
  \institution{The Hong Kong Polytechnic University}
  \city{Hong Kong}
  \country{China}
}

\author{Qixin Wang}
\email{qixin.wang@polyu.edu.hk}
\affiliation{%
  \institution{The Hong Kong Polytechnic University}
  \city{Hong Kong}
  \country{China}
}
\authornote{Corresponding author: Qixin Wang (qixin.wang@polyu.edu.hk)}

\author{Yu Pei}
\email{max.yu.pei@polyu.edu.hk}
\affiliation{%
  \institution{The Hong Kong Polytechnic University}
  \city{Hong Kong}
  \country{China}
}

\renewcommand{\shortauthors}{Shiyu et al.}

\begin{abstract}
Reinforcement Learning (RL) has gained significant attention across various domains. However, the increasing complexity of RL programs presents testing challenges, particularly the oracle problem: defining the correctness of the RL program. Conventional human oracles struggle to cope with the complexity, leading to inefficiencies and potential unreliability in RL testing. To alleviate this problem, we propose an automated oracle approach that leverages RL properties using fuzzy logic. Our oracle quantifies an agent's behavioral compliance with reward policies and analyzes its trend over training episodes. It labels an RL program as "Buggy" if the compliance trend violates expectations derived from RL characteristics. We evaluate our oracle on RL programs with varying complexities and compare it with human oracles. Results show that while human oracles perform well in simpler testing scenarios, our fuzzy oracle demonstrates superior performance in complex environments. The proposed approach shows promise in addressing the oracle problem for RL testing, particularly in complex cases where manual testing falls short. It offers a potential solution to improve the efficiency, reliability, and scalability of RL program testing. This research takes a step towards automated testing of RL programs and highlights the potential of fuzzy logic-based oracles in tackling the oracle problem.

\end{abstract}

\ccsdesc[300]{Software and its engineering~Software testing and debugging}

\keywords{Reinforcement Learning, Test Oracle, Fuzzy Logic, Reward Function Variation}

\maketitle

\section{Introduction}
Nowadays, \emph{data-driven AI machine learning} (referred to as ``\emph{AI machine learning}'' in the following for simplicity) is being rapidly adopted across various domains~\cite{jordan2015machine}\cite{zhang2020machine}. However, as the complexities of AI machine learning programs increase, they encounter growing testing challenges\cite{ribeiro2020beyond}\cite{braiek2020testing}.

Unlike conventional human testing, as the complexities of nowadays programs soar, test automation is inevitable. One critical tool in the workflow of automatic program testing is \emph{oracle}. Given a set of test cases and program execution outputs, an oracle is the tool that automatically labels the correctness of the execution outputs (hence labels the correctness of the execution). It is well-known that oracles can be extremely difficult to design, depending on the application domain. In fact, for many application domains, we do not yet have good enough oracles, hence ``human oracles'' are still the main stream~\cite{barr2014oracle}, blocking the full automation of program testing. This is the so-called \emph{oracle problem}~\cite{barr2014oracle}. 

For the application domain of AI machine learning, the oracle problem is even bigger. It faces a unique challenge: how to define whether the output of a machine learning program is correct or wrong. For example, an AI machine learning program for facial recognition outputs a \emph{Neural Network} (NN). How to judge whether this NN is correct or wrong? All that we can do is to throw a test set (a collection of pictures involving human faces and other contents) to the NN, and the NN outputs a set of recognition decisions. Even if there are wrong decision (recognizing a human as car, or the reverse), it does not mean the AI learning program is buggy; and vice versa.

\begin{figure}[htpb]
\centering
\includegraphics[width=\linewidth]{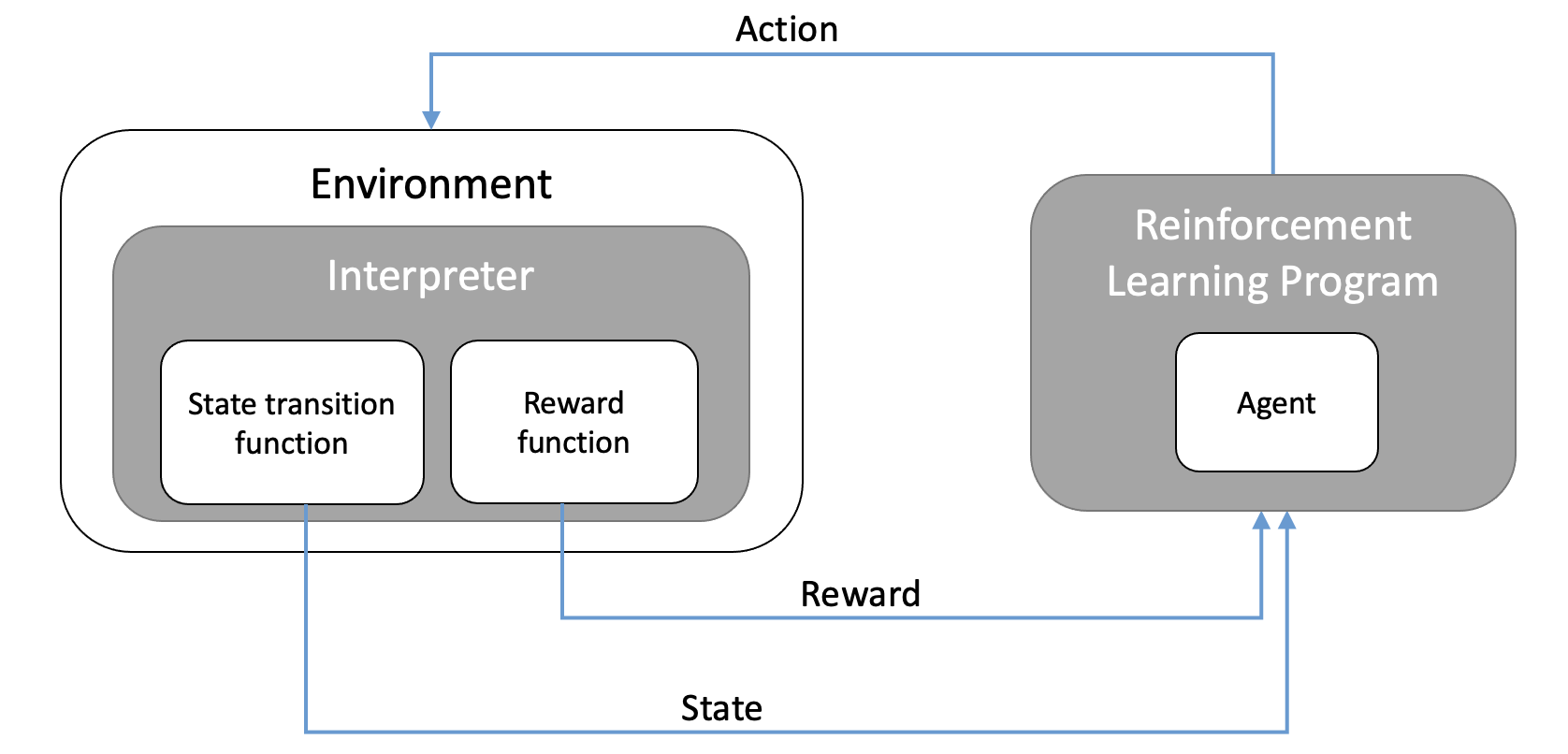}
\caption{Reinforcement Learning Framework}
\label{fig:Reinforcement_Learning}
\end{figure}

To address this problem for generic AI machine learning is too big a challenge. In this paper, we are going to focus on \emph{Reinforcement Learning} (RL)~\cite{sutton2018reinforcement}\cite{watkins1992q}\cite{singh2000convergence}. RL is a promising sub-category of AI machine learning~\cite{li2017deep}\cite{moerland2023model}. As illustrated by Fig.~\ref{fig:Reinforcement_Learning}, an RL program trains an \emph{agent} (e.g. a NN to autopilot a car) to interact (aka ``\emph{game}'') with an \emph{environment} (e.g. the actual physical environment of the car auto-piloting, or more commonly, the simulator of the physical environment). Each action applied by the agent to the environment will change the state of the environment (e.g. the position of the car), and trigger a reward. The agent is supposed to maximize the cumulative rewards~\cite{kaelbling1996reinforcement}. This naturally matches the underlying demands of a wide range of applications that pursue rewards, such as video-game~\cite{lample2017playing},  recommendation systems~\cite{zheng2018drn}, and robotics (optimal control) ~\cite{kober2013reinforcement}. In turn, the increasing demands for RL also foster more complex RL programs. The growing complexities of the RL programs demand test automation, hence testing oracles. As a main-stream sub-category of AI machine learning, the oracle problem on AI machine learning also stands out in RL.

Unfortunately, the RL oracle problem is understudied. To our best knowledge, human oracles are still the main-stream for testing RL programs. To better automate the testing, in this paper, we are going to propose a non-human oracle, which exploits the features of RL. Specifically, we design our oracle upon the following two heuristics.

\vspace{0.3in}

\noindent \textbf{Heuristic 1: Behavioral Compliance to Reward Policies}

One feature of RL is that the environment always contains a \emph{reward function} (see Fig.~\ref{fig:Reinforcement_Learning}), which rewards ``good'' actions and punishes ``bad'' actions depending on the current state of the environment. We call such state-dependent preferences the ``\emph{policy}'' of the reward function. If an RL program is bugless, it should respect the policy: throughout an \emph{episode} of training (an episode is a complete time-series of interactions between the agent and the environment, e.g. playing a video-game from the ``game start'' to the ``game over''), it should make the agent's actions (i.e. \emph{behavior}) increasingly compliant to the policy~\cite{sutton2018reinforcement}\cite{watkins1992q}\cite{singh2000convergence}.

Based on this feature, we propose to randomly generate a set of $I$ policies via the well-known \emph{reward function variation}~\cite{kirk2023survey} method. For each policy, we carry out $E$ episodes of RL training of the agent. In each episode, we use fuzzy logic~\cite{berenji1992learning}\cite{hullermeier2005fuzzy} to quantify each action's compliance to the given policy. The resulted policy-compliance-value time-series should show an increasing trend throughout the episode. If not, the RL program is suspicious. Our oracle shall label it as ``Buggy.''

\vspace{0.1in}

\noindent \textbf{Heuristic 2: Behavioral Abnormality after Convergence}

The increasing policy-compliance-value time-series mentioned in \textbf{Heuristic 1} may or may not converge in an episode of training. However, if it converges, an RL program normally should not drastically damage this convergence (specifically, drastically reduce the policy-compliance-value for several consecutive steps)~\cite{bertsekas1996neuro}\cite{gordon1996chattering}. 

We also exploit this feature as part of our oracle. If an RL program drastically reduce the policy-compliance-value
in an increasing and converging policy-compliance-value time-series for enough many consecutive steps, we regard the RL program suspicious. Our oracle shall label it as ``Buggy.''

\vspace{0.1in}

Guided by the above heuristics, this paper makes the following contributions:

\textbf{1)} We propose an oracle that exploits the features of RL for their testing.

\textbf{2)} We compare our oracle with the conventional human oracle, the result shows that our oracle can 
outperform the human oracle under some test-beds, and achieve comparable performance under other test-beds. This means, when the human oracle is unavailable, people can use our proposed oracle to test RL programs.

\vspace{0.1in}

The rest of our paper is structured as follows:
Section~\ref{sect:Background} provides background on RL and fuzzy logic. Section~\ref{sect:Solution} presents our proposed oracle. Section~\ref{sect:Test Oracle for RL} describes the experimental setup and results, demonstrating the effectiveness of our approach. Section V lists some related works and explains the differences between our work and others. Finally, Section VI discusses future research directions and concludes the paper.

The source code is accessible at our GitHub repository \cite{RLtestingnew} \href{https://github.com/QixinWangCpsLab/RL-testing-new}{\textcolor{blue}{RL-testing-new}}.





\section{Background}
\label{sect:Background}

\subsection{Fuzzy Logic}
\label{subsect:FuzzyLogic}

Fuzzy logic is a method designed for dealing with uncertainty. It allows computers to understand and manipulate concepts that are imprecise or vague. It was introduced by Lotfi Zadeh in 1965 and was designed to overcome the limitations of traditional binary logic (true or false) when dealing with fuzzy concepts\cite{zadeh1965fuzzy}. 

\begin{figure}
    \centering
    \includegraphics[width=\linewidth]{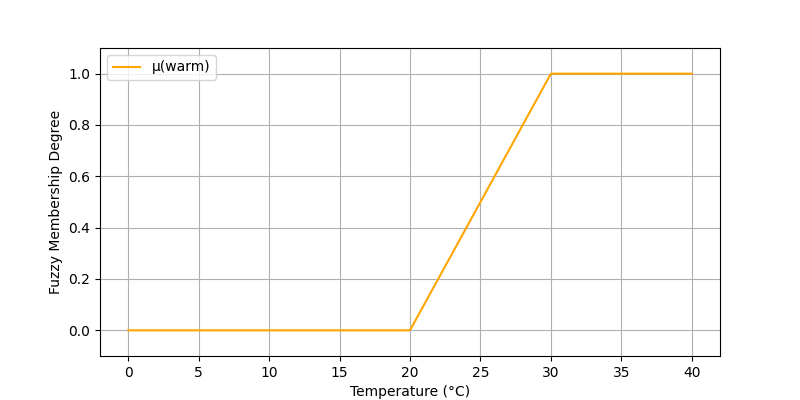}
    \caption{Fuzzy Membership Function of "Warm"}
    \label{fig:Fuzzy_Menbership}
\end{figure}

Fuzzy membership is a core concept within fuzzy logic that quantifies the degree of truth or the likelihood of a particular element being part of a fuzzy set. Unlike classical set theory, where an element either belongs or does not belong to a set, fuzzy membership allows for a spectrum of membership. This is expressed through a membership function, typically denoted as $\mu$
, which assigns each element a membership value between 0 and 1, where 0 represents complete non-membership and 1 represents full membership. 
The values in between reflect partial membership, providing a mathematical way to capture the tiny differences of linguistic terms such as "warm", "slow" or "high". Fig~\ref{fig:Fuzzy_Menbership} is an example of judging "warm" using fuzzy membership function.

Fuzzy membership functions are critical for applications that simulate human-like reasoning, enabling systems to make decisions with imprecise or incomplete information\cite{klir1996fuzzy}.

\subsection{Reinforcement Learning}

In the context of reinforcement learning, we have the following components\cite{sutton2018reinforcement}:

\begin{itemize}
  \item \textbf{Environment:} The external system with which the agent interacts, typically denoted by $\mathcal{E}$. It is defined by a set of states $S$ and a set of actions $A$ available to the agent, and it governs the state transitions and rewards.
  
  \item \textbf{Interpreter:} A function within the environment that interprets the actions taken by the agent and returns a new state and reward. It can be seen as the combination of the state transition function and the reward function.
  
  \item \textbf{State:} A representation of the current situation that the agent is in, denoted by $s \in S$. Each state provides the necessary information for the agent to make decisions.
  
  \item \textbf{Reward:} A scalar feedback signal, denoted by $r \in \mathbb{R}$, which indicates the immediate return of transitioning from state $s$ to state $s'$ due to action $a$. The reward function is defined as $R: S \times A \rightarrow \mathbb{R}$.
  
  \item \textbf{Agent:} An entity that learns from the environment through interactions. It aims to select actions that maximize cumulative rewards. The agent is defined by its policy $\pi$, where $\pi: S \rightarrow A$.
  
  \item \textbf{Action:} A decision made by the agent that results in a transition from one state to another within the environment. An action is denoted by $a \in A$.

\end{itemize}

The interaction between the agent and the environment in a reinforcement learning system can be described as follows:

\begin{equation}
(s_t, a_t, r_{t}, s_{t+1}) \sim P(\cdot | s_t, a_t)
\end{equation}

where $s_t$ is the state at time $t$, $a_t$ is the action taken by the agent, $r_{t}$ is the reward received after taking action $a_t$, and $s_{t+1}$ is the new state after the transition. $P$ represents the probability of the transition and reward given the current state and action.

The agent's objective is to learn a policy $\pi^*$ that maximizes the expected cumulative reward described in reward function\cite{beggs2005convergence}, which can be represented as:

\begin{equation}
\pi^* = \arg\max_{\pi} \mathbb{E}\left[\sum_{t=0}^{\infty} \gamma^t R(s_t, \pi(s_t)) | \pi \right]
\end{equation}

where $\gamma \in [0,1]$ is the discount factor, balancing immediate and future rewards.

\section{Solution}
\label{sect:Solution}

%

\subsection{Proposed Solution}
\label{subsect:ProposedSolution}

\subsubsection{Intended Policy}
\label{subsubsect:IntendedPolicy}

In our setup, we define a reward function which is guided by a `intended policy'. The `intended policy' maps a set of selected states to the corresponding ideal actions, represented by $\pi^\star : S^{\star} \mapsto A^\star$, where $S^{\star} \subseteq S$ and $A^\star \subseteq A$.

The determination of `intended policies' can either be accomplished in a stochastic approach or by designing $S^{\star}$ and $A^\star$ based on specific needs. Within the scope of this research, the design of `intended policies' was performed using the stochastic approach.

\subsubsection{Fuzzy Reward Function}
To enhance the adaptability of our oracle in various environments(e.g. environment with continuous or discrete action and state spaces) and facilitate the learning process for RL program, we employ fuzzy logic to design a robust and learnable reward function. This approach enables RL program to gradually learn the 'intended policy' and we can use this approach to evaluate the policy-compliance-value which is a metric for measuring the degree of compliance of the learned policy in the RL program with respect to the 'intended policy'. We use this policy-compliance-value metric in the design of our oracle.

In order to calculate policy-compliance-value, we need to calculate step-compliance-value and it is determined by state-compliance-value and action-compliance-value. Both the state-compliance-value and action-compliance-value serve as indicators of fuzzy membership. For simplicity, each state $s$ will be assigned to a unique and closest reference state $s^\star \in S^\star$. The distance function $d_{\text{state}}: S \times S \rightarrow \mathbb{R}$ can be defined based on the environment. The minimum distance $\delta$ among the reference states in $S^\star$ is:

\begin{equation}
\delta = \min_{\substack{s_i^\star, s_j^\star \in S^\star \\ s_i^\star \neq s_j^\star}} d_{\text{state}}(s_i^\star, s_j^\star)
\end{equation}

The state-compliance-value $\mu_{\text{state}}(s, s^\star) \in [0, 1]$ for state $s$ with respect to $s^\star$ is:

\begin{equation}
\mu_{\text{state}}(s, s^\star) =
\begin{cases}
0, & \text{if } d_{\text{state}}(s, s^\star) > \frac{1}{2} \delta \\
f_{\text{state}}(d_{\text{state}}(s, s^\star)), & \text{otherwise}
\end{cases}
\end{equation}

Here, $f_{\text{state}} \in [0, 1]$ and it is a non-increasing fuzzy membership function, with a higher value indicating greater fuzzy membership degree and vice versa.

For any $s \in S$, we use $s^\triangle \in S^\star$ to denote the closest reference state with the current state $s$, that is:
\begin{equation}
s^\triangle = \arg\min_{s^\star \in S^\star} d_{\text{state}}(s, s^\star)
\end{equation}

To determine action-compliance-value $\mu_{\text{action}}$, we consider the intended policy action $a^\triangle \in A^\star$, where $a^\triangle=\pi^{\star}$($s^\triangle$), along with the actual action $a$. Then, $\mu_{\text{action}}(a, a^\triangle)$ is defined as:

\begin{equation}
\mu_{\text{action}}(a, a^\triangle) = f_{\text{action}}(d_{\text{action}}(a, a^\triangle))
\end{equation}

where $d_{\text{action}}: A \times A \rightarrow \mathbb{R}$ is the action distance metric, and $f_{\text{action}} \in [0, 1]$ is a non-increasing fuzzy membership function. $d_{\text{action}}$ can be defined according to user requirements.

The step-compliance-value, which a indicator of fuzzy membership, $\mu_{\text{step}}$, is a composite function of state-compliance-value and action-compliance-value:

\begin{equation}
\label{eq:myfuzzy}
\mu_{\text{step}}(s, a) = \mu_{\text{state}}(s, s^\triangle) \times \mu_{\text{action}}(a, a^\triangle)
\end{equation}

We can now represent the compliance degree of the current state-action pair with the state-action pair guided by 'intended policy' through measuring $\mu_{\text{step}}(s, a)$ using Equation~\ref{eq:myfuzzy}. In order to relate the state-action pair to the RL program's performance while maintaining a positive correlation with the mentioned fuzzy compliance metrics, we define fuzzy reward function $R_{\text{step}}(s, a)$, for any state-action pair $(s, a)$, the resulting value is positively correlated with $\mu_{\text{step}}(s, a)$ from Equation~\ref{eq:myfuzzy}. If $\mu_{\text{step}}(s, a)$ increases, then $R_{\text{step}}(s, a)$ is also expected to increase, reinforcing the compliance with the intended policy.

\begin{equation}
R_{\text{step}}(s, a) \propto \mu_{\text{step}}(s, a)
\label{eq:positiveCorrelationR}
\end{equation}

Here, \( \propto \) denotes a positive correlation between the output of the function $R_{\text{step}}(s, a)$ and the step-compliance-value $\mu_{\text{step}}(s, a)$. The exact functional form of $R_{\text{step}}$ is not specified and can be modeled to satisfy the specific characteristics and performance requirements of the system.

Finally, in each epoch $e$, we can calculate policy-compliance-value $\mu_e$ of the learned policy with respect to the intended policy using the log list $L_e$ containing (state, action) traces in the training process and a policy-compliance-value threshold $\theta_{\textsf{poli-cmpl}}$:

\begin{equation}
\mu_e = \frac{\sum\limits_{\substack{(state, action) \in L_e \\ \mu_{step}(state, action) \geqslant \theta_{\textsf{poli-cmpl}}}} \mu_{step}(state, action)} {|\forall (state, action) \in L_e, \mu_{step}(state, action) \geqslant \theta_{\textsf{poli-cmpl}}|}
\label{eq:policy-compliance-value}
\end{equation}

The policy-compliance-value $\mu_e$ is an indicator of fuzzy membership between learned policy and intended policy. It is a metric to evaluate the functionality of an RL program. A high policy-compliance-value indicates a strong compliance of the RL program to the intended policy, reflecting good functionality. Conversely, a low policy-compliance-value suggests a deviation from the intended policy, indicating potential issues with the program's functionality.

\subsubsection{Test Oracle for RL}
\label{sect:Test Oracle for RL}


\textbf{Algorithm \ref{alg:rl_oracle}} presents the oracle main function for testing reinforcement learning programs. In the initialization phase, the algorithm generates a set of intended policies $\Pi^{\star}$ and their corresponding reward functions $\mathcal{R}$. Each intended policy $\pi_i^{\star}$ is a mapping from selected states set $S_i^{\star}$ to to the corresponding ideal actions set $A_i^{\star}$.

In the $i_{th}$ epoch, the algorithm executes the reinforcement learning program $\mathcal{P}$ with reward function $R_i$. Our algorithm will record the $(\text{state}, \text{action})$ traces of the execution in a log list $\mathcal{L}_{i}$. We will repeat this process for $E$ times.

In the log analysis phase, the algorithm calculates the policy-compliance-value time-series $\alpha$ of the learned policy represented by the log list $\mathcal{L}$,with respect to intended policy $\pi^{\star}$, using the $CalcPolicyComplianceValueTimeSeries$ function(presented in Algorithm \ref{alg:rl_oracle_pt2}) with a policy-compliance-value threshold $\theta_{\text{poli-cmpl}}$. It then analyzes the trend of the policy-compliance-value time-series $\alpha$ using the $TrendAnalysis$ function (presented in Algorithm \ref{alg:rl_oracle_pt3}), which checks learning trend of the learned policy and behavioral abnormality after time-series $\alpha$ converges.

Finally, in the oracle phase, the algorithm determines the buggy or non-buggy nature of the reinforcement learning program based on the proportion of 'True' values in the set $\mathcal{T}$, which contains result of trend analysis for each intended policy. If the proportion is greater than or equal to a threshold $\theta_{orcl}$, the program is regarded as non-buggy; otherwise, it is regarded as buggy.

Our oracle main function provides a systematic approach to testing reinforcement learning programs by evaluating their learning progress against a set of intended policies. By analyzing the trends of policy-compliance-values of the learned policies with respect to the intended policies, and checking behavioral abnormality after convergence, the algorithm can effectively detect potential bugs and assess the quality of the reinforcement learning program. The check for behavioral abnormality after convergence adds an additional layer of analysis, providing a more comprehensive assessment of the program's performance.

%
\begin{algorithm}[H]
\caption{Oracle Main Function for Testing Reinforcement Learning Programs}
\label{alg:rl_oracle}
\setcounter{ctr}{0}
\begin{tabbing}
8888\=888\=888\=888\=888\=888\=888\=888\=888\=888\=888\kill
\stp{} \textbf{function} OracleMain ( \\
\> $\mathcal{P}$: the to-be-tested reinforcement learning program, \\
\> $\mathcal{E}$: the environment, \\
\> $I$: the number of intended policies to generate, \\
\> $E$: the number of training epochs for each intended policy, \\
\> $\theta_{\textsf{orcl}} \in [0, 1]$: the threshold constant used directly by the oracle, \\
\> $n \in \{1, 2, \ldots, E-1\}$: sliding-window size for judging \\
\>\> policy-compliance-value time-series' convergence or \\
\>\> convergence-abnormality, \\
\> $\varepsilon \in \mathbb{R}^+$: the threshold to judge the convergence of a \\
\>\> policy-compliance-value time-series, \\
\> $\Delta \in \mathbb{R}^+$: the threshold to judge the abnormalities in a \\
\>\> converging policy-compliance-value time-series, \\
\> $\theta_{\textsf{poli-cmpl}} \in [0, 1]$: the threshold to judge if a (game) step \\
\>\> in an epoch complies with the intended policy): \\
       \> \\
\stp{} \> // Initialization Phase: \\
\stp{} \> Generate a set of intended policies $\Pi^{\star} = \{\pi_i^{\star}\}_{i=1, \ldots, I}$ and \\
       \> the corresponding reward functions $\mathcal{R} = \{R_i\}_{i=1, \ldots, I}$;\\
\stp{} \> // An intended policy $\pi_i^{\star}$ is a function $S_i^{\star} \mapsto A_i^{\star}$, where \\
\stp{} \> // $S_i^{\star} \subseteq S$ is a set of selected states and $A_i^{\star} \subseteq A$ is a set\\
\stp{} \> // of corresponding ideal actions.
\\
\stp{} \> \\
\stp{} \> // Training Phase: \\
\stp{} \> \textbf{for} ($i: 1 \leqslant i \leqslant I$) \{ \\
\stp{} \>\> Initialize the log list $\mathcal{L}_{i} \leftarrow []$;\\
\stp{} \>\> \textbf{for} ($e: 1 \leqslant e \leqslant E$) \{ \\
\stp{} \>\>\> // The $e$th training epoch: \\
\stp{} \>\>\> $L_{i, e} \leftarrow$ execute the RL program $\mathcal{P}$\\
       \>\>\> with the environment $\mathcal{E}$ and the reward function $R_i$; \\
\stp{} \>\>\> // $L_{i, e}$ is a list of (state, action)s, representing the \\
\stp{} \>\>\> // step-by-step trace of the above execution, note the \\
\stp{} \>\>\> // length of $L_{i, e}$ is decided by the game or epoch  \\
\stp{} \>\>\> // ending condition. \\
\stp{} \>\>\> Append $L_{i, e}$ as an element to the list $\mathcal{L}_{i}$; \\
\stp{} \>\> \} \\
\stp{} \> \} \\
\stp{} \> \\
\stp{} \> // Log Analysis Phase: \\
\stp{} \> Initialize the trends value set $\mathcal{T} \leftarrow \varnothing$; \\
\stp{} \> \textbf{for} ($i: 1 \leqslant i\leqslant I$) \{ \\ 
\stp{} \>\> List $\alpha_i \leftarrow $ CalcPolicyComplianceValueTimeSeries( \\
       \>\>\> $\pi_i^{\star}$, $\mathcal{L}_i$, $\theta_{\textsf{poli-cmpl}}$); \\
\stp{} \>\>  Boolean $\tau_i \leftarrow $ TrendAnalysis($\alpha_i$, n, $\varepsilon$, $\Delta$); \\
%
\stp{} \>\> Add $\tau_i$ as an element to the set $\mathcal{T}$; \\
\stp{} \> \} \\
       \> \\
\stp{} \> // Oracle Phase: \\
\stp{} \> $c \leftarrow $ number of ``True'' values in $\mathcal{T}$; \\
\stp{} \> \textbf{if} $c/|\mathcal{T}| \geqslant \theta_{\textsf{orcl}}$ \{ \textbf{return} Non-buggy; \} \textbf{else} \{ \textbf{return} Buggy; \}
\end{tabbing}
\end{algorithm}
\normalsize

As presented in \textbf{Algorithm \ref{alg:rl_oracle_pt2}} , the $ CalcPolicyComplianceValue-
TimeSeries$ function is a subroutine of the oracle main function. It calculates the policy-compliance-value time-series of the learned policy with respect to the intended policy$\pi^\star$. This function takes three inputs: the intended policy $\pi^{\star}$, a list of logs $\mathcal{L}$ recording the (state, action) traces during the training process across $E$ epochs, and a policy-compliance-value threshold $\theta_{\text{poli-cmpl}}$. It initializes an empty list $\alpha$ to store the policy-compliance-value time-series for each epoch. For each epoch $e$, the function retrieves the list of $(\text{state}, \text{action})$ traces $L_e$ from the log list $\mathcal{L}$. It then iterates through each $(\text{state}, \text{action})$ trace in $L_e$ and calculates the step-compliance-value of each $(\text{state}, \text{action})$ trace using $\pi^{\star}$.

\begin{algorithm}[H]
\caption{Oracle Subroutine 1: the CalcPolicyComplianceValueTimeSeries Function}
\label{alg:rl_oracle_pt2}
\setcounter{ctr}{0}
\begin{tabbing}
8888\=888\=888\=888\=888\=888\=888\=888\=888\=888\=888\kill
\stp{} \textbf{function} CalcPolicyComplianceValueTimeSeries( \\
       \> $\pi^{\star}$: the intended policy (i.e. a function $S^{\star} \mapsto A^{\star}$, where $S^{\star} \subseteq S$ \\
       \>\> is a set of selected states), and $A_i^{\star} \subseteq A$ is a set of correspond-\\
       \>\> ing ideal actions\\
       \> $\mathcal{L}$: list of $E$ epochs of logs of training process, \\
       \> $\theta_{\textsf{poli-cmpl}}$: a threshold constant where $0 \leqslant \theta_{\textsf{poli-cmpl}} \leqslant 1$ ): \\
       \> \\
\stp{} \> Initialize the policy-compliance-value time-series $\alpha \leftarrow []$; \\ 
\stp{} \> \textbf{for} ($e: 1 \leqslant e \leqslant $ length of $\mathcal{L}$  (i.e. $E$)) \{ \\
\stp{} \>\> $L_e \leftarrow $ the $e$th element of $\mathcal{L}$; \\
\stp{} \>\> // $L_e$ is a list of $(\text{state}, \text{action})$ pairs, representing the \\
       \>\> // $(\text{state}, \text{action})$ traces in the $e$th epoch of training process. \\
\stp{} \>\> policy-compliance-value (of the epoch's learned policy \\
       \>\> with respect to $\pi^{\star}$) $\mu_e \leftarrow 0$; \\
\stp{} \>\> $c_e \leftarrow 0$; \\
\stp{} \>\> \textbf{for} ($j: 1 \leqslant j \leqslant$ length of $L_e$) \{\\
\stp{} \>\>\> $(\text{state}_j, \text{action}_j) \leftarrow$ the $j$th element of $L_e$; \\
\stp{} \>\>\> $\text{state}_j^{\star} \leftarrow$ the closest state in $S^{\star}$ to $\text{state}_j$; \\
\stp{} \>\>\> $\text{action}_j^{\star} \leftarrow \pi^{\star}(\text{state}_j^{\star})$; \\
\stp{} \>\>\> State-compliance-value $\mu_{\text{state}} \leftarrow $ fuzzy state-compliance\\
\>\>\>\> -value between $\text{state}_j$ and $\text{state}_j^{\star}$; \\
\stp{} \>\>\> Action-compliance-value $\mu_{\text{action}} \leftarrow $ fuzzy action-compli-\\
       \>\>\>\> ance-value between $\text{action}_j$ and $\text{action}_j^{\star}$; \\
\stp{} \>\>\> Step-compliance-value $\mu_{\text{step}} \leftarrow \mu_{\text{state}} \times \mu_{\text{action}}$; \\
\stp{} \>\>\> \textbf{if} ($\mu_{\text{state}} \geqslant \theta_{\textsf{poli-cmpl}}$) \\
\stp{} \>\>\>\> $\mu_e \leftarrow \mu_e + \mu_{\text{step}}$; \\
\stp{} \>\>\>\> $c_e \leftarrow c_e + 1$; \\
\stp{} \>\>\> \} \\
\stp{} \>\> \}  \\
\stp{} \>\> \textbf{if} $c_e > 0$ \{ \\
\stp{} \>\>\> $\mu_e \leftarrow \mu_e / c_e$; \\
\stp{} \>\>\> Append $\mu_e$ as an element to the list $\alpha$; \\
\stp{} \>\> \} \textbf{else} \{\\
\stp{} \>\>\> Append $0$ to the list of $\alpha$; \\
\stp{} \>\> \} \\
\stp{} \> \} \\
\stp{} \> \textbf{return} $\alpha$;
\end{tabbing}

\end{algorithm}

The state-compliance-value $\mu_{\text{state}}$ is computed as the fuzzy membership between the given state $\text{s}_j$ and its closest reference state $\text{s}j^{\triangle} \in S^{\star}$. Similarly, the action-compliance-value $\mu_{\text{action}}$ is calculated as the fuzzy membership between the given action $\text{a}_j$ and the intended action $\text{a}_j^{\triangle}$ corresponding to $\text{s}_j^{\triangle}$.

The step-compliance-value $\mu_{\text{step}}$ is then computed as the product of $\mu_{\text{state}}$ and $\mu_{\text{action}}$. If the step-compliance-value $\mu_{\text{step}}$ is greater than or equal to the threshold $\theta_{\text{poli-cmpl}}$, the step-compliance-value is added to the epoch's policy-compliance-value $\mu_e$, and the counter $c_e$ is incremented by 1.

After iterating through all the $(\text{state}, \text{action})$ traces in the epoch, if the counter $c_e$ is greater than zero, the epoch's policy-compliance-value $\mu_e$ is calculated as $\mu_e$ divided by $c_e$ and appended to the policy-compliance-value list $\alpha$. If $c_e$ is zero, a value of zero is appended to $\alpha$ instead.

Finally, the function returns the list of policy-compliance-values $\alpha$, where each element represents the fuzzy membership between the learned policy and the intended policy of each training epoch.

The $CalcPolicyComplianceValueTimeSeries$ function provides an approach to quantify the fuzzy membership between the learned policies and the intended policy using fuzzy logic. By considering state-compliance-value, action-compliance-value and a policy-compliance-value threshold, it captures the degree to which the reinforcement learning program complies with the desired behavior defined by the intended policy.

\begin{algorithm}[H]
\caption{Oracle Subroutine 3: the TrendAnalysis Function}
\label{alg:rl_oracle_pt3}
\setcounter{ctr}{0}
\begin{tabbing}
8888\=888\=888\=888\=888\=888\=888\=888\=888\=888\=888\kill
\stp{} \textbf{function} TrendAnalysis( \\
       \> $\alpha_i$: a policy-compliance-value time-series, \\
       \> $n \in \mathbb{N}^+$: sliding-window size for analyzing $\alpha_i$, \\
       \> $\varepsilon \in \mathbb{R}^+$: the threshold to judge the convergence of a \\
       \>\> policy-compliance-value time-series, \\
\stp{} \> slope $\leftarrow$ LinearRegression($\alpha_i$); \\
\stp{} \> cnvg\_idx $\leftarrow$ TimeSeriesConvergenceStartingIndex($\alpha_i$, $n$, $\varepsilon$); \\
\stp{} \> \textbf{if} (slope $< 0$) \{ \textbf{return} False; \} \\
\stp{} \> \textbf{if} (slope $\geqslant 0$ and cnvg\_idx does not exist) \{ \textbf{return} True; \} \\
\stp{} \> // now slope $\geqslant 0$ and cnvg\_idx exists (i.e. $\alpha_i$ converges) \\
%
\stp{} \> // find max and min difference in post-convergence interval.\\
\stp{} \> interval\_max\_min\_diff $=$ \\
       \>\> MaxMinDiff($\alpha_{j}[$cnvg\_idx : cnvg\_idx + n$]$) \\
\stp{} \> // calculate lower bound for detecting behavioral abnormality\\
\stp{} \> lower\_bound$=\alpha_{j}[$cnvg\_idx$]-$interval\_max\_min\_diff\\
\stp{} \> // search for $n$ consecutive violations of the lower\_bound \\
\stp{} \> c $\leftarrow 0$; \\
\stp{} \> \textbf{for} ($j$: $($cnvg\_idx$+1) \leqslant j \leqslant |\alpha_i|)$) \{ \\
\stp{} \>\> \textbf{if} ($\alpha_i[j] <$ lower\_bound ) \{ c $\leftarrow$ c $+ 1$; \} \textbf{else} \{ c $\leftarrow 0$; \} \\
\stp{} \>\> \textbf{if} (c $\geqslant n$) \{ \textbf{return} False; \} \\
\stp{} \> \} \\
\stp{} \> \textbf{return} True; \\
\end{tabbing}
\end{algorithm}

\textbf{Algorithm \ref{alg:rl_oracle_pt3}} analyzes the trend of  policy-compliance-values $\alpha_i$. It first calculates the linear regression slope of the policy-compliance-values in $\alpha_i$. If the slope is negative, we return "False" to indicate that this program is buggy. Otherwise, we will then check behavioral abnormality after Convergence. The method of determining convergence needs to be chosen based on the user requirements. In our program, we assume, if there are $n$ consecutive points fluctuating in a small interval $\epsilon$, it indicates convergence, where $n$ is sliding window size for analyzing $\alpha_i$. If $\alpha_i$ doesn't converge, we will ignore the behavioral abnormality check and return "True". In the abnormality check, we initially calculate a lower bound by subtracting the value at the convergence point from the maximum-minimum difference within an array of size $n$, taken from the elements following the convergence point in $\alpha_i$. Then it checks behavioral abnormality after the convergence point. If there are more than $n$ consecutive points lower than the calculated lower bound, we will return "False" to indicate this program is buggy. Otherwise, we will return "True" to indicate this program is non-buggy.


\section{Evaluation}
\label{sect:Evaluation}
\subsection{Experiment Setup}

\subsubsection{Testbed}
Our experiments are based on Stable Baselines3 (SB3) \cite{stable-baselines3}. SB3 is a widely-used open-source library for reinforcement learning (RL) that provides reliable implementations of several state-of-the-art RL algorithms. Built on top of PyTorch, SB3 supports various RL algorithms, including Deep Q-Networks (DQN), Proximal Policy Optimization (PPO), Soft Actor-Critic (SAC), and more.

In our experiments, we focus on testing three RL algorithms: DQN, A2C, and PPO. We evaluate these algorithms on two gymnasium environments\cite{towers_gymnasium_2023}:

\noindent\textbf{Frozenlake}: A simple game where the agent navigates in a environment to reach a goal state while avoiding falling into holes. We test all three algorithms (DQN, A2C, and PPO) on this environment.

\noindent\textbf{MountainCar Continuous}: A classic control problem where a car must climb a steep hill by building momentum. Since DQN is not supported in the MountainCar environment, we only test A2C and PPO on this environment.

\subsubsection{Bugs.}
To simulate scenarios where errors may occur in the implementation of RL algorithms, we inject bugs into the SB3 program. These bugs can be divided into four categories: \textbf{training}, \textbf{model}, \textbf{updating network}, and \textbf{exploring the environment}. The \textbf{training} category involves bugs encountered during the training process, such as loss calculation and hyperparameter selection. The \textbf{model} category includes bugs related to the model itself, such as model structure and initialization. Bugs associated with updating the network are classified as \textbf{updating network}, such as updating the q-network in the DQN algorithm. During the training process, the agent prioritizes selecting actions with high-value rewards that have been tried before, but also needs to explore actions that have not been chosen previously, which is known as exploration. Any bugs related to this process are categorized as \textbf{exploring the environment}. We designed 39 bugs for different algorithms and formed 56 bug versions of SB3 by injecting them into different algorithms on different environments.

\subsubsection{Baseline.}
To validate the effectiveness of our method, we set up a control group consisting of a Human Expert team to serve as a Human Oracle. The Human Expert team was carefully selected to ensure a high level of expertise and diverse backgrounds. It included one member with a Ph.D. degree, two graduate students with extensive development experience, and two undergraduate students with research experience in relevant fields. We recreated conditions that closely resemble real-world RL program development for the Human Oracle team. For each combination of environment, algorithm, and bug, we first record an animation of the untrained agent's behavior in the environment. After training is complete, we record another animation showcasing the agent's behavior in the environment post-training. We also use Tensorboard\cite{tensorflow2015-whitepaper}, which is a widely used tool in NN developing, to track changes in reward and loss throughout the training process. The Human Expert team combines the pre- and post-training animations with the reward and loss curves provided by Tensorboard to determine whether the agent's training process is successful, thereby assessing if the Reinforcement Learning Program contains any bugs.

We then used the buggy versions of the SB3 program on the selected environments and algorithms using our method described in Section~\ref{sect:Test Oracle for RL}. The effectiveness of our testing method in detecting the injected bugs is measured using metrics like false negative rate, false positive rate, Accuracy, Precision, Recall and F1 scores. We also compared the performance of our oracle (denoted as Fuzzy Oracle in this chapter) with the performance of the Human Oracle.

\subsubsection{Research Questions}
In our research, the following research questions are studied:

\noindent\textbf{RQ1:} How should the parameters and thresholds of Fuzzy Oracle be set?


\noindent\textbf{RQ2:} How does the Fuzzy Oracle compare to the Human Oracle in different environments and algorithms??

By addressing these research questions, we aim to provide a comprehensive evaluation of our proposed testing method for reinforcement learning programs, considering various aspects such as parameter setting, effectiveness across different scenarios, impact of environment complexity, and comparison with human oracle.


\subsection{Evaluation and Results}
\subsubsection{RQ1: How should the parameters and thresholds of our test oracle be set?}

\begin{figure}[htbp]
\centering
\includegraphics[width=0.45\textwidth]{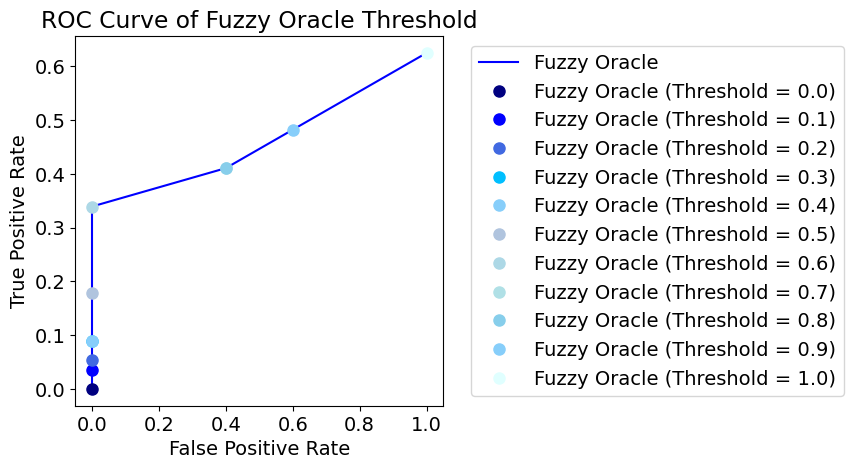}
\caption{ROC curve for the Fuzzy Oracle at different threshold values}
\label{fig:roc_curve}
\end{figure}

The selection of parameters in Algorithm~\ref{alg:rl_oracle} plays a crucial role in determining the experimental results. Our qualitative analysis reveals several key insights.

First, due to the limited computational resources, we chose to generate a modest number of 10 intended policies, denoted by $I$. This choice strikes a balance between exploring different scenarios in the environment and managing computational costs.

Second, when training in the Frozenlake environment, we set the number of training epochs $E$ to 300, with each epoch consisting of 200 training steps. For the more complex and challenging Mountaincar continuous environment, we increased the number of epochs to 700 to ensure sufficient training.

Third, the parameters related to the convergence break detection, namely the number of consecutive values $n$ and the threshold $\theta_{\text{break}}$, need to be determined based on the specific circumstances. In our experiments, we set $n=5$ and $\Delta=0.1$, meaning that a convergence break is identified if the fuzzy similarity with the intended policy drops by more than 10\% for 5 consecutive epochs after convergence.

Fourth, the choice of the state similarity threshold $\theta_{\textsf{poli-cmpl}}$ depends on the specific calculation method for fuzzy similarities. In this paper, we set a more relaxed fuzzy threshold with the intention of ensuring a lower false positive rate.

Regarding the selection of the threshold $\theta$, we conducted a thorough analysis using the ROC curve, as shown in Figure~\ref{fig:roc_curve}. Based on the ROC curve, we decided to set the threshold value to 0.7 for our experiments. This choice allows us to maintain a relatively low false positive rate (FPR) while achieving a satisfactory true positive rate (TPR). By selecting a threshold of 0.7, we strike a balance between minimizing false positives and obtaining good bug detection performance.

In practice, the selection of these parameters requires consideration of problem-specific factors, computational resource constraints, and performance requirements. Fine-tuning and adapting the parameters based on the specific problem can enhance the accuracy of assessing reinforcement learning programs and effectively identify bugs.

\subsubsection{RQ2: How does the Fuzzy Oracle compare to the Human Oracle in different environments and algorithms?}

\begin{figure}[htbp]
\centering
\includegraphics[width=0.45\textwidth]{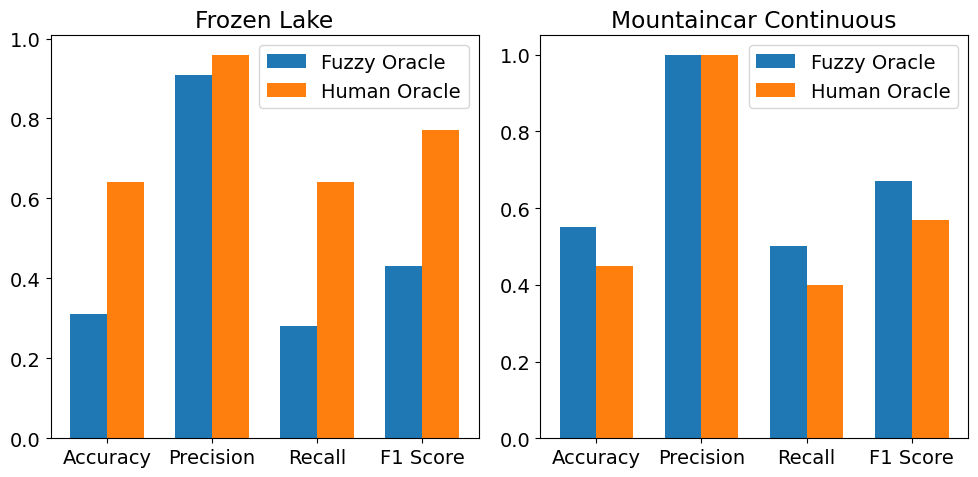}
\caption{Fuzzy Oracle vs. Human Oracle in different Environments}
\label{fig:different_env}
\end{figure}

\begin{figure}[htbp]
\centering
\includegraphics[width=0.45\textwidth]{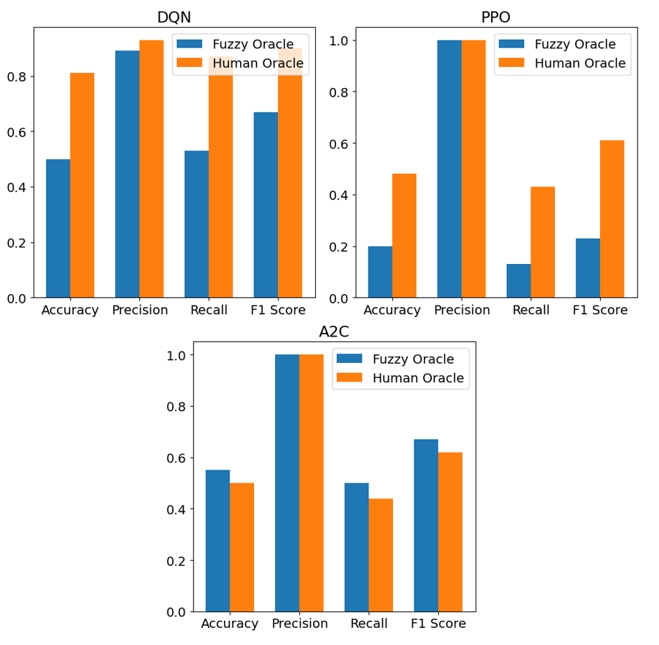}
\caption{Fuzzy Oracle vs. Human Oracle in different Algorithms}
\label{fig:different_algorithms}
\end{figure}

To address this research question, we analyzed the performance of the Fuzzy Oracle and the Human Oracle across different environments (Frozen Lake and Mountaincar Continuous) and algorithms (A2C, DQN, and PPO). The results showed differences in their performance based on these factors, as shown in Table \ref{tab:performance_oracle_environment}, Table \ref{tab:performance_algorithm_oracle}, Fig. \ref{fig:different_env}, and Fig. \ref{fig:different_algorithms}.

In terms of environmental differences, the Fuzzy Oracle exhibited a higher false negative rate in the Frozen Lake environment, missing 27 positive instances. In contrast, its false negative rate was significantly lower in the Mountaincar Continuous environment, with only 10 positive instances being missed (see Table \ref{tab:performance_oracle_environment}). Compared to the Human Oracle, the Fuzzy Oracle achieved a higher precision (0.91) in the Frozen Lake environment, indicating its ability to minimize false positives. However, its accuracy (0.31), recall (0.28), and F1 score (0.43) were lower than the Human Oracle (see Fig. \ref{fig:different_env}). In the Mountaincar Continuous environment, the Fuzzy Oracle showed improved performance with an accuracy of 0.55, precision of 1.00, recall of 0.50, and an F1 score of 0.67, suggesting that it is more adept at making judgments in complex environments compared to the Human Oracle.

\begin{table}[h]
\centering
\begin{tabular}{l|l|cccc}
\hline
\textbf{Oracle} & \textbf{Testbed} & \textbf{TP} & \textbf{FP} & \textbf{TN} & \textbf{FN} \\
\hline
\multirow{2}{*}{Fuzzy Oracle} & Frozen Lake & 9 & 0 & 3 & 27 \\
& Mountaincar Continuous & 10 & 0 & 2 & 10 \\
\hline
\multirow{2}{*}{Human Oracle} & Frozen Lake & 23 & 1 & 2 & 13 \\
& Mountaincar Continuous & 8 & 0 & 2 & 12 \\
\hline
\end{tabular}
\caption{\small Performance by Oracle and Environment}
\label{tab:performance_oracle_environment}
\end{table}

When examining the Oracles' performance across different algorithms (Table \ref{tab:performance_algorithm_oracle} and Fig. \ref{fig:different_algorithms}), we observed that the Fuzzy Oracle's false negative rate was highest for the PPO algorithm, with 20 positive instances being missed. For data generated by the A2C algorithm, it missed 9 positive instances, while for the DQN algorithm, it missed 8 positive instances. The Fuzzy Oracle achieved the highest accuracy (0.55), precision (1.00), recall (0.50), and F1 score (0.67) with the A2C algorithm, indicating its effectiveness in detecting bugs in this specific context. However, its performance with DQN and PPO was lower compared to the Human Oracle.

\begin{table}[h]
\centering
\begin{tabular}{l|l|cccc}
\hline
\textbf{Oracle} & \textbf{Algorithm} & \textbf{TP} & \textbf{FP} & \textbf{TN} & \textbf{FN} \\
\hline
\multirow{3}{*}{Fuzzy Oracle} & A2C & 9 & 0 & 2 & 9 \\
& DQN & 7 & 0 & 1 & 8 \\
& PPO & 3 & 0 & 2 & 20 \\
\hline
\multirow{3}{*}{Human Oracle} & A2C & 8 & 0 & 2 & 10 \\
& DQN & 13 & 1 & 0 & 2 \\
& PPO & 10 & 0 & 2 & 13 \\
\hline
\end{tabular}
\caption{\small Performance by Algorithm and Oracle}
\label{tab:performance_algorithm_oracle}
\end{table}

The experimental results demonstrate that the Fuzzy Oracle's performance differs from the Human Oracle across various environments and algorithms. These differences can be attributed to the specific characteristics of the environments and the nature of the algorithms. The Fuzzy Oracle's performance appears to be influenced by the environment in which it operates, as evidenced by the lower false negative rate in the Mountaincar Continuous environment compared to the Frozen Lake environment. Additionally, the choice of algorithm seems to impact the Fuzzy Oracle's performance, with the A2C algorithm resulting in the highest accuracy, precision, recall, and F1 score among the three algorithms tested.

These findings highlight the importance of considering the impact of environment and algorithm selection when evaluating and applying Oracles. The Fuzzy Oracle demonstrates its superiority in handling complex environments, showcasing its robustness and adaptability in such contexts. However, its performance with different algorithms varies, and further research is needed to improve its adaptability across different algorithmic settings. Future research should focus on enhancing the Fuzzy Oracle's performance and comparing it with the Human Oracle in a wider range of environments and algorithms to gain a more comprehensive understanding of their strengths and limitations.






\section{Related Work}
\label{sect:RelatedWork}
\subsection{Fuzzy Logic Testing}
Fuzzy logic has been applied in various aspects of software testing to handle uncertainty and provide more comprehensive evaluations. Researchers have explored the use of fuzzy logic in test case generation, test case prioritization, and test oracle construction.

In the area of test case generation, Bai et al. \cite{bai2008automatically} proposed a fuzzy logic-based approach for generating test cases from software specifications. They used fuzzy set theory to model the input domain and employed fuzzy inference rules to determine the most effective test cases. This approach allowed for the generation of test cases that cover the input space more comprehensively and efficiently compared to traditional crisp methods. However, their work focuses on generating test cases for general software systems, while our approach specifically targets the testing of reinforcement learning programs.

Fuzzy logic has also been used in test case prioritization to improve the effectiveness of testing efforts. Xu et al. \cite{xu2010method} developed a fuzzy logic-based approach for prioritizing test cases in regression testing. They considered various factors, such as code coverage, fault detection history, and execution time, and used fuzzy inference rules to assign priority values to test cases. This approach helped in identifying the most critical test cases and optimizing the testing process. In contrast, our work focuses on evaluating the learning progress of reinforcement learning programs rather than prioritizing test cases.

Wan et al. propose DRLFuzz\cite{wan2024coverage}, a coverage-guided fuzzing framework for testing Deep Reinforcement Learning systems, focusing on generating diverse failure cases. While sharing the common goal of improving RL system reliability, our research differs by addressing challenges arising from the complexity and diversity of RL implementations. We introduce fuzzy logic for test oracle design and propose reward function variation to expand test coverage, while DRLFuzz employs coverage-guided fuzzing to automatically generate test cases. Despite the differences in approach, both studies tackle challenges like complexity and uncertainty in RL testing, and the ideas presented in DRLFuzz offer valuable insights for our research.

Test oracle construction is another area where fuzzy logic has been applied. Shahamiri et al. \cite{shahamiri2011automated} proposed a fuzzy logic-based approach for constructing test oracles in the absence of explicit specifications. They used fuzzy inference rules to determine the expected behavior of the system based on its inputs and outputs. This approach provided a more flexible and robust way of constructing test oracles, especially in scenarios where precise specifications are not available. While their work deals with test oracle construction for general software systems, our approach specifically uses fuzzy logic to compare learned policies with intended policies in the context of reinforcement learning.


\subsection{Reinforcement Learning Testing and Debugging}
Reward-function variation has the potential to become a benchmark in the area of zero-shot Generalization of RL, yet it is very under-studied\cite{kirk2023survey}. Similarly, we have not observed this promising technique being utilized within the test oracle designing of Reinforcement Learning. Nonetheless, we have identified some related work in the testing of RL:

DRLinter\cite{nikanjam2022faults} proposed by Guo et al., is a rule-based static analysis tool for detecting potential faults in DRL programs. DRLinter defines a set of rules based on expert knowledge and best practices to identify common antipatterns and bad practices, such as suboptimal reward shaping, incorrect tensor operations, and inappropriate hyperparameter settings. In contrast, our work focuses on the dynamic testing of RL programs by combining fuzzy logic and reward function variation to create a more robust test oracle. We aim to handle the ambiguity and uncertainty inherent in reward functions and explore a wide range of scenarios through reward function variations. While DRLinter and our research address different aspects of DRL system reliability, they complement each other in the pursuit of developing robust and dependable DRL applications, highlighting the importance of combining static analysis and dynamic testing techniques for comprehensive quality assurance.

STARLA\cite{zolfagharian2023search} is a search-based testing approach for DRL agents that aims to find faulty execution episodes efficiently. It uses genetic algorithms to search the state-action space for sequences that are likely to trigger faults, guided by machine learning models trained on historical episode data to predict the likelihood of faults. STARLA also incorporates state abstraction techniques to reduce the search space by clustering similar states based on the agent's perception, i.e., the optimal state-action value function. STARLA determines whether the model has triggered faults by identifying "faulty states". In STARLA, a faulty state is defined as a state that violates predefined requirements and is often an end state (termination state). For example, in an autonomous driving scenario, a state where the vehicle collides with an obstacle is considered a faulty state. However, STARLA differs from our proposed oracle method in its focus and granularity of analysis. While STARLA emphasizes detecting faults that occur in specific states within an episode, our method focuses on analyzing the agent's learning process and progress over time by providing a continuous assessment of the agent's policy similarity trend during training. Combining STARLA's fine-grained, state-level fault detection with our continuous, learning-oriented analysis could potentially lead to a more comprehensive and explainable DRL testing approach, enabling us to identify faults in specific episodes and gain insights into the agent's evolving behavior throughout the training process.

MDP Playground\cite{rajan2023mdp} provides a way to analyze and debug RL agents by controlling various dimensions of hardness, allowing for experiments on toy environments and studying the effects on complex environments. It aims to improve agent performance and adaptability by examining factors such as non-Markov information states, irrelevant features, representations, and low-level dimensions of hardness. The difference between \cite{rajan2023mdp} and our approach is that MDP Playground provided a testbed, thus simplifies the creation of test environments for RL algorithms, but it still uses test set and evaluation metrics as RL test oracle. In contrast, Reward-function variation is a more universal and straightforward which does not rely on a test set or evaluation metrics.

Mohammad et al. \cite{wardat2021deeplocalize} provides us a white-box testing approach for Deep Neural Networks. This paper introduces a method for detecting errors in deep neural networks by tracking and analyzing the flow of values during training, using imperative representations and probes for dynamic analysis to pinpoint faulty layers or hyperparameters, and an algorithm to identify root causes by monitoring numerical errors and the influence of each component on the model's performance. There are two main differences between this work and our reward-variation method: Firstly, RL algorithms are not always Deep Neural Networks, for example, Q-learning doesn’t rely on Deep Neural Network. In that case, \cite{wardat2021deeplocalize} may not work. Secondly, our Reward-function variation method treats the model as black box while \cite{wardat2021deeplocalize} needs to dig into the model. Therefore, the two methods are suitable for different scenarios.

Maximilian et al. \cite{igl2019generalization} introduces a novel regularization technique for RL algorithms—Selective Noise Injection (SNI) and Information Bottleneck (IB)—which can significantly enhance the generalization of agents in new environments. The SNI and IB methods are used during the training phase to enhance the generalization abilities of the model, while our method is used during the testing phase.

Gu et al. \cite{gu2022muffin} proposed an approach for testing deep learning libraries by creating varied neural network structures to uncover bugs, focusing on ensuring the libraries' reliability for tasks including training and inference phases. Muffin assesses foundational software integrity, our method works as test oracle for RL algorithms.


\section{Conclusion}
\label{sect:Conclusion}
In this paper, we addressed the oracle problem in testing Reinforcement Learning (RL) programs by proposing a novel automated testing approach using fuzzy logic. Our proposed oracle exploits the inherent features of RL, namely the behavioral compliance to reward policies and the behavioral abnormality after convergence. By quantifying an agent's compliance with the intended policy using fuzzy logic and analyzing its trend over training episodes, our oracle can effectively identify suspicious behaviors and label them as potentially buggy.

We conducted experiments to evaluate our proposed oracle and compared its performance with a human oracle across different RL environments and algorithms. The results showed that the performance of our oracle varies depending on the complexity of the environment and the choice of algorithm. In simpler environments like Frozen Lake, the human oracle outperformed our fuzzy oracle. However, in more complex environments such as Mountaincar Continuous, our oracle demonstrated superior performance, highlighting its ability to handle intricate scenarios effectively.

The experimental results also revealed that the choice of algorithm influences the performance of our oracle. It achieved the highest accuracy, precision, recall, and F1 score with the A2C algorithm, indicating its effectiveness in detecting bugs in this specific context. However, its performance with DQN and PPO was lower compared to the human oracle, suggesting the need for further research to improve its adaptability across different algorithmic settings.

Our research takes a significant step towards automating the testing of RL programs by addressing the oracle problem. The proposed fuzzy logic-based oracle provides a promising solution for detecting bugs and assessing the correctness of RL agents' behaviors, especially in complex environments where manual testing becomes challenging. However, there is still room for improvement in terms of the oracle's adaptability to different algorithms and its performance in simpler environments.

Future research directions include refining the fuzzy logic-based approach to enhance its robustness and generalizability across various RL settings. Additionally, exploring the integration of other techniques, such as metamorphic testing or property-based testing, could further improve the effectiveness of automated testing for RL programs.

In conclusion, our work contributes to the field of automated testing for RL programs by proposing a novel fuzzy logic-based oracle approach. While the results demonstrate its potential, particularly in complex environments, further research is needed to address its limitations and improve its adaptability. As RL continues to gain popularity across various domains, developing reliable and efficient automated testing techniques becomes increasingly crucial to ensure the quality and reliability of RL applications.

\section{Acknowledgments}
We would like to express our sincere gratitude to Mr. Pan Zewen, Mr. Haryoto Dylan, Mr. Xiang Yufan, Mr. Zhang Yibin, and Dr. Shen Henghua for their valuable contributions as the Human Oracle in this research. Their expertise and insights were instrumental in validating the effectiveness of our proposed method. We are truly grateful for their support and collaboration.

The research project related to this paper was also supported in part by the Hong Kong Research Grants Council (RGC) Theme-based Research Scheme T22-505/19-N (P0031331, RBCR, P0031259, RBCP); 
in part by RGC Germany/HK Joint Research Scheme G-PolyU503/16; 
in part by HKSAR Government and HKJCCT P0041424 (ZB5A); 
and in part by the Hong Kong Polytechnic University P0042701 (CE09), P0046487 (CE0F), P0047916 (TACW), P0042699 (CE55), P0045578 (CE1C), P0043884 (CD6R), P0047965 (TAEB), P0047964 (TAEA), P0033695 (ZVRD), P0013879 (BBWH), P0036469 (CDA8), P0043634 (1-TAB2), P0043647 (1-TABF), P0042721 (1-ZVG0),
LTG22-25/IICA/33 (49NZ), and TDG22-25/SMS-11 (49RF).

\bibliographystyle{ACM-Reference-Format}
\bibliography{ISSTA2024}


\begin{thebibliography}{36}


\ifx \showCODEN    \undefined \def \showCODEN     #1{\unskip}     \fi
\ifx \showDOI      \undefined \def \showDOI       #1{#1}\fi
\ifx \showISBNx    \undefined \def \showISBNx     #1{\unskip}     \fi
\ifx \showISBNxiii \undefined \def \showISBNxiii  #1{\unskip}     \fi
\ifx \showISSN     \undefined \def \showISSN      #1{\unskip}     \fi
\ifx \showLCCN     \undefined \def \showLCCN      #1{\unskip}     \fi
\ifx \shownote     \undefined \def \shownote      #1{#1}          \fi
\ifx \showarticletitle \undefined \def \showarticletitle #1{#1}   \fi
\ifx \showURL      \undefined \def \showURL       {\relax}        \fi
\providecommand\bibfield[2]{#2}
\providecommand\bibinfo[2]{#2}
\providecommand\natexlab[1]{#1}
\providecommand\showeprint[2][]{arXiv:#2}

\bibitem[Abadi et~al\mbox{.}(2015)]%
        {tensorflow2015-whitepaper}
\bibfield{author}{\bibinfo{person}{Mart\'{i}n Abadi}, \bibinfo{person}{Ashish Agarwal}, \bibinfo{person}{Paul Barham}, \bibinfo{person}{Eugene Brevdo}, \bibinfo{person}{Zhifeng Chen}, \bibinfo{person}{Craig Citro}, \bibinfo{person}{Greg~S. Corrado}, \bibinfo{person}{Andy Davis}, \bibinfo{person}{Jeffrey Dean}, \bibinfo{person}{Matthieu Devin}, \bibinfo{person}{Sanjay Ghemawat}, \bibinfo{person}{Ian Goodfellow}, \bibinfo{person}{Andrew Harp}, \bibinfo{person}{Geoffrey Irving}, \bibinfo{person}{Michael Isard}, \bibinfo{person}{Yangqing Jia}, \bibinfo{person}{Rafal Jozefowicz}, \bibinfo{person}{Lukasz Kaiser}, \bibinfo{person}{Manjunath Kudlur}, \bibinfo{person}{Josh Levenberg}, \bibinfo{person}{Dandelion Man\'{e}}, \bibinfo{person}{Rajat Monga}, \bibinfo{person}{Sherry Moore}, \bibinfo{person}{Derek Murray}, \bibinfo{person}{Chris Olah}, \bibinfo{person}{Mike Schuster}, \bibinfo{person}{Jonathon Shlens}, \bibinfo{person}{Benoit Steiner}, \bibinfo{person}{Ilya Sutskever}, \bibinfo{person}{Kunal Talwar},
  \bibinfo{person}{Paul Tucker}, \bibinfo{person}{Vincent Vanhoucke}, \bibinfo{person}{Vijay Vasudevan}, \bibinfo{person}{Fernanda Vi\'{e}gas}, \bibinfo{person}{Oriol Vinyals}, \bibinfo{person}{Pete Warden}, \bibinfo{person}{Martin Wattenberg}, \bibinfo{person}{Martin Wicke}, \bibinfo{person}{Yuan Yu}, {and} \bibinfo{person}{Xiaoqiang Zheng}.} \bibinfo{year}{2015}\natexlab{}.
\newblock \bibinfo{title}{{TensorFlow}: Large-Scale Machine Learning on Heterogeneous Systems}.
\newblock
\newblock
\urldef\tempurl%
\url{https://www.tensorflow.org/}
\showURL{%
\tempurl}
\newblock
\shownote{Software available from tensorflow.org}.


\bibitem[Bai and Chen(2008)]%
        {bai2008automatically}
\bibfield{author}{\bibinfo{person}{Shih-Ming Bai} {and} \bibinfo{person}{Shyi-Ming Chen}.} \bibinfo{year}{2008}\natexlab{}.
\newblock \showarticletitle{Automatically constructing grade membership functions of fuzzy rules for students’ evaluation}.
\newblock \bibinfo{journal}{\emph{Expert Systems with Applications}} \bibinfo{volume}{35}, \bibinfo{number}{3} (\bibinfo{year}{2008}), \bibinfo{pages}{1408--1414}.
\newblock


\bibitem[Barr et~al\mbox{.}(2014)]%
        {barr2014oracle}
\bibfield{author}{\bibinfo{person}{Earl~T Barr}, \bibinfo{person}{Mark Harman}, \bibinfo{person}{Phil McMinn}, \bibinfo{person}{Muzammil Shahbaz}, {and} \bibinfo{person}{Shin Yoo}.} \bibinfo{year}{2014}\natexlab{}.
\newblock \showarticletitle{The oracle problem in software testing: A survey}.
\newblock \bibinfo{journal}{\emph{IEEE transactions on software engineering}} \bibinfo{volume}{41}, \bibinfo{number}{5} (\bibinfo{year}{2014}), \bibinfo{pages}{507--525}.
\newblock


\bibitem[Beggs(2005)]%
        {beggs2005convergence}
\bibfield{author}{\bibinfo{person}{Alan~W Beggs}.} \bibinfo{year}{2005}\natexlab{}.
\newblock \showarticletitle{On the convergence of reinforcement learning}.
\newblock \bibinfo{journal}{\emph{Journal of economic theory}} \bibinfo{volume}{122}, \bibinfo{number}{1} (\bibinfo{year}{2005}), \bibinfo{pages}{1--36}.
\newblock


\bibitem[Berenji and Khedkar(1992)]%
        {berenji1992learning}
\bibfield{author}{\bibinfo{person}{Hamid~R Berenji} {and} \bibinfo{person}{Pratap Khedkar}.} \bibinfo{year}{1992}\natexlab{}.
\newblock \bibinfo{booktitle}{\emph{Learning and tuning fuzzy logic controllers through reinforcements}}.
\newblock \bibinfo{type}{{T}echnical {R}eport}.
\newblock


\bibitem[Bertsekas and Tsitsiklis(1996)]%
        {bertsekas1996neuro}
\bibfield{author}{\bibinfo{person}{Dimitri Bertsekas} {and} \bibinfo{person}{John~N Tsitsiklis}.} \bibinfo{year}{1996}\natexlab{}.
\newblock \bibinfo{booktitle}{\emph{Neuro-dynamic programming}}.
\newblock \bibinfo{publisher}{Athena Scientific}.
\newblock


\bibitem[Braiek and Khomh(2020)]%
        {braiek2020testing}
\bibfield{author}{\bibinfo{person}{Houssem~Ben Braiek} {and} \bibinfo{person}{Foutse Khomh}.} \bibinfo{year}{2020}\natexlab{}.
\newblock \showarticletitle{On testing machine learning programs}.
\newblock \bibinfo{journal}{\emph{Journal of Systems and Software}}  \bibinfo{volume}{164} (\bibinfo{year}{2020}), \bibinfo{pages}{110542}.
\newblock


\bibitem[Gordon(1996)]%
        {gordon1996chattering}
\bibfield{author}{\bibinfo{person}{Geoffrey~J Gordon}.} \bibinfo{year}{1996}\natexlab{}.
\newblock \bibinfo{booktitle}{\emph{Chattering in SARSA (lambda)-a CMU learning lab internal report}}.
\newblock \bibinfo{type}{{T}echnical {R}eport}. \bibinfo{institution}{Technical report). Carnegie Mellon University}.
\newblock


\bibitem[Gu et~al\mbox{.}(2022)]%
        {gu2022muffin}
\bibfield{author}{\bibinfo{person}{Jiazhen Gu}, \bibinfo{person}{Xuchuan Luo}, \bibinfo{person}{Yangfan Zhou}, {and} \bibinfo{person}{Xin Wang}.} \bibinfo{year}{2022}\natexlab{}.
\newblock \showarticletitle{Muffin: Testing deep learning libraries via neural architecture fuzzing}. In \bibinfo{booktitle}{\emph{Proceedings of the 44th International Conference on Software Engineering}}. \bibinfo{pages}{1418--1430}.
\newblock


\bibitem[H{\"u}llermeier(2005)]%
        {hullermeier2005fuzzy}
\bibfield{author}{\bibinfo{person}{Eyke H{\"u}llermeier}.} \bibinfo{year}{2005}\natexlab{}.
\newblock \showarticletitle{Fuzzy methods in machine learning and data mining: Status and prospects}.
\newblock \bibinfo{journal}{\emph{Fuzzy sets and Systems}} \bibinfo{volume}{156}, \bibinfo{number}{3} (\bibinfo{year}{2005}), \bibinfo{pages}{387--406}.
\newblock


\bibitem[Igl et~al\mbox{.}(2019)]%
        {igl2019generalization}
\bibfield{author}{\bibinfo{person}{Maximilian Igl}, \bibinfo{person}{Kamil Ciosek}, \bibinfo{person}{Yingzhen Li}, \bibinfo{person}{Sebastian Tschiatschek}, \bibinfo{person}{Cheng Zhang}, \bibinfo{person}{Sam Devlin}, {and} \bibinfo{person}{Katja Hofmann}.} \bibinfo{year}{2019}\natexlab{}.
\newblock \showarticletitle{Generalization in reinforcement learning with selective noise injection and information bottleneck}.
\newblock \bibinfo{journal}{\emph{Advances in neural information processing systems}}  \bibinfo{volume}{32} (\bibinfo{year}{2019}).
\newblock


\bibitem[Jordan and Mitchell(2015)]%
        {jordan2015machine}
\bibfield{author}{\bibinfo{person}{Michael~I Jordan} {and} \bibinfo{person}{Tom~M Mitchell}.} \bibinfo{year}{2015}\natexlab{}.
\newblock \showarticletitle{Machine learning: Trends, perspectives, and prospects}.
\newblock \bibinfo{journal}{\emph{Science}} \bibinfo{volume}{349}, \bibinfo{number}{6245} (\bibinfo{year}{2015}), \bibinfo{pages}{255--260}.
\newblock


\bibitem[Kaelbling et~al\mbox{.}(1996)]%
        {kaelbling1996reinforcement}
\bibfield{author}{\bibinfo{person}{Leslie~Pack Kaelbling}, \bibinfo{person}{Michael~L Littman}, {and} \bibinfo{person}{Andrew~W Moore}.} \bibinfo{year}{1996}\natexlab{}.
\newblock \showarticletitle{Reinforcement learning: A survey}.
\newblock \bibinfo{journal}{\emph{Journal of artificial intelligence research}}  \bibinfo{volume}{4} (\bibinfo{year}{1996}), \bibinfo{pages}{237--285}.
\newblock


\bibitem[Kirk et~al\mbox{.}(2023)]%
        {kirk2023survey}
\bibfield{author}{\bibinfo{person}{Robert Kirk}, \bibinfo{person}{Amy Zhang}, \bibinfo{person}{Edward Grefenstette}, {and} \bibinfo{person}{Tim Rockt{\"a}schel}.} \bibinfo{year}{2023}\natexlab{}.
\newblock \showarticletitle{A survey of zero-shot generalisation in deep reinforcement learning}.
\newblock \bibinfo{journal}{\emph{Journal of Artificial Intelligence Research}}  \bibinfo{volume}{76} (\bibinfo{year}{2023}), \bibinfo{pages}{201--264}.
\newblock


\bibitem[Klir and Yuan(1996)]%
        {klir1996fuzzy}
\bibfield{author}{\bibinfo{person}{George~J Klir} {and} \bibinfo{person}{Bo Yuan}.} \bibinfo{year}{1996}\natexlab{}.
\newblock \showarticletitle{Fuzzy sets and fuzzy logic: theory and applications}.
\newblock \bibinfo{journal}{\emph{Possibility Theory versus Probab. Theory}} \bibinfo{volume}{32}, \bibinfo{number}{2} (\bibinfo{year}{1996}), \bibinfo{pages}{207--208}.
\newblock


\bibitem[Kober et~al\mbox{.}(2013)]%
        {kober2013reinforcement}
\bibfield{author}{\bibinfo{person}{Jens Kober}, \bibinfo{person}{J~Andrew Bagnell}, {and} \bibinfo{person}{Jan Peters}.} \bibinfo{year}{2013}\natexlab{}.
\newblock \showarticletitle{Reinforcement learning in robotics: A survey}.
\newblock \bibinfo{journal}{\emph{The International Journal of Robotics Research}} \bibinfo{volume}{32}, \bibinfo{number}{11} (\bibinfo{year}{2013}), \bibinfo{pages}{1238--1274}.
\newblock


\bibitem[Lample and Chaplot(2017)]%
        {lample2017playing}
\bibfield{author}{\bibinfo{person}{Guillaume Lample} {and} \bibinfo{person}{Devendra~Singh Chaplot}.} \bibinfo{year}{2017}\natexlab{}.
\newblock \showarticletitle{Playing FPS games with deep reinforcement learning}. In \bibinfo{booktitle}{\emph{Proceedings of the AAAI Conference on Artificial Intelligence}}, Vol.~\bibinfo{volume}{31}.
\newblock


\bibitem[Li(2017)]%
        {li2017deep}
\bibfield{author}{\bibinfo{person}{Yuxi Li}.} \bibinfo{year}{2017}\natexlab{}.
\newblock \showarticletitle{Deep reinforcement learning: An overview}.
\newblock \bibinfo{journal}{\emph{arXiv preprint arXiv:1701.07274}} (\bibinfo{year}{2017}).
\newblock


\bibitem[Moerland et~al\mbox{.}(2023)]%
        {moerland2023model}
\bibfield{author}{\bibinfo{person}{Thomas~M Moerland}, \bibinfo{person}{Joost Broekens}, \bibinfo{person}{Aske Plaat}, \bibinfo{person}{Catholijn~M Jonker}, {et~al\mbox{.}}} \bibinfo{year}{2023}\natexlab{}.
\newblock \showarticletitle{Model-based reinforcement learning: A survey}.
\newblock \bibinfo{journal}{\emph{Foundations and Trends{\textregistered} in Machine Learning}} \bibinfo{volume}{16}, \bibinfo{number}{1} (\bibinfo{year}{2023}), \bibinfo{pages}{1--118}.
\newblock


\bibitem[Nikanjam et~al\mbox{.}(2022)]%
        {nikanjam2022faults}
\bibfield{author}{\bibinfo{person}{Amin Nikanjam}, \bibinfo{person}{Mohammad~Mehdi Morovati}, \bibinfo{person}{Foutse Khomh}, {and} \bibinfo{person}{Houssem Ben~Braiek}.} \bibinfo{year}{2022}\natexlab{}.
\newblock \showarticletitle{Faults in deep reinforcement learning programs: a taxonomy and a detection approach}.
\newblock \bibinfo{journal}{\emph{Automated software engineering}} \bibinfo{volume}{29}, \bibinfo{number}{1} (\bibinfo{year}{2022}), \bibinfo{pages}{8}.
\newblock


\bibitem[QixinWangCpsLab(2024)]%
        {RLtestingnew}
\bibfield{author}{\bibinfo{person}{QixinWangCpsLab}.} \bibinfo{year}{2024}\natexlab{}.
\newblock \bibinfo{title}{{RL-testing-new: PyTorch version of Stable Baselines, reliable implementations of reinforcement learning algorithms}}.
\newblock \bibinfo{howpublished}{\url{https://github.com/QixinWangCpsLab/RL-testing-new}}.
\newblock


\bibitem[Raffin et~al\mbox{.}(2021)]%
        {stable-baselines3}
\bibfield{author}{\bibinfo{person}{Antonin Raffin}, \bibinfo{person}{Ashley Hill}, \bibinfo{person}{Adam Gleave}, \bibinfo{person}{Anssi Kanervisto}, \bibinfo{person}{Maximilian Ernestus}, {and} \bibinfo{person}{Noah Dormann}.} \bibinfo{year}{2021}\natexlab{}.
\newblock \showarticletitle{Stable-Baselines3: Reliable Reinforcement Learning Implementations}.
\newblock \bibinfo{journal}{\emph{Journal of Machine Learning Research}} \bibinfo{volume}{22}, \bibinfo{number}{268} (\bibinfo{year}{2021}), \bibinfo{pages}{1--8}.
\newblock
\urldef\tempurl%
\url{http://jmlr.org/papers/v22/20-1364.html}
\showURL{%
\tempurl}


\bibitem[Rajan et~al\mbox{.}(2023)]%
        {rajan2023mdp}
\bibfield{author}{\bibinfo{person}{Raghu Rajan}, \bibinfo{person}{Jessica Lizeth~Borja Diaz}, \bibinfo{person}{Suresh Guttikonda}, \bibinfo{person}{Fabio Ferreira}, \bibinfo{person}{Andr{\'e} Biedenkapp}, \bibinfo{person}{Jan~Ole von Hartz}, {and} \bibinfo{person}{Frank Hutter}.} \bibinfo{year}{2023}\natexlab{}.
\newblock \showarticletitle{MDP Playground: An Analysis and Debug Testbed for Reinforcement Learning}.
\newblock \bibinfo{journal}{\emph{Journal of Artificial Intelligence Research}}  \bibinfo{volume}{77} (\bibinfo{year}{2023}), \bibinfo{pages}{821--890}.
\newblock


\bibitem[Ribeiro et~al\mbox{.}(2020)]%
        {ribeiro2020beyond}
\bibfield{author}{\bibinfo{person}{Marco~Tulio Ribeiro}, \bibinfo{person}{Tongshuang Wu}, \bibinfo{person}{Carlos Guestrin}, {and} \bibinfo{person}{Sameer Singh}.} \bibinfo{year}{2020}\natexlab{}.
\newblock \showarticletitle{Beyond accuracy: Behavioral testing of NLP models with CheckList}.
\newblock \bibinfo{journal}{\emph{arXiv preprint arXiv:2005.04118}} (\bibinfo{year}{2020}).
\newblock


\bibitem[Shahamiri et~al\mbox{.}(2011)]%
        {shahamiri2011automated}
\bibfield{author}{\bibinfo{person}{Seyed~Reza Shahamiri}, \bibinfo{person}{Wan Mohd Nasir~Wan Kadir}, \bibinfo{person}{Suhaimi Ibrahim}, {and} \bibinfo{person}{Siti Zaiton~Mohd Hashim}.} \bibinfo{year}{2011}\natexlab{}.
\newblock \showarticletitle{An automated framework for software test oracle}.
\newblock \bibinfo{journal}{\emph{Information and Software Technology}} \bibinfo{volume}{53}, \bibinfo{number}{7} (\bibinfo{year}{2011}), \bibinfo{pages}{774--788}.
\newblock


\bibitem[Singh et~al\mbox{.}(2000)]%
        {singh2000convergence}
\bibfield{author}{\bibinfo{person}{Satinder Singh}, \bibinfo{person}{Tommi Jaakkola}, \bibinfo{person}{Michael~L Littman}, {and} \bibinfo{person}{Csaba Szepesv{\'a}ri}.} \bibinfo{year}{2000}\natexlab{}.
\newblock \showarticletitle{Convergence results for single-step on-policy reinforcement-learning algorithms}.
\newblock \bibinfo{journal}{\emph{Machine learning}}  \bibinfo{volume}{38} (\bibinfo{year}{2000}), \bibinfo{pages}{287--308}.
\newblock


\bibitem[Sutton and Barto(2018)]%
        {sutton2018reinforcement}
\bibfield{author}{\bibinfo{person}{Richard~S Sutton} {and} \bibinfo{person}{Andrew~G Barto}.} \bibinfo{year}{2018}\natexlab{}.
\newblock \bibinfo{booktitle}{\emph{Reinforcement Learning: An Introduction}}.
\newblock \bibinfo{publisher}{MIT press}.
\newblock


\bibitem[Towers et~al\mbox{.}(2023)]%
        {towers_gymnasium_2023}
\bibfield{author}{\bibinfo{person}{Mark Towers}, \bibinfo{person}{Jordan~K. Terry}, \bibinfo{person}{Ariel Kwiatkowski}, \bibinfo{person}{John~U. Balis}, \bibinfo{person}{Gianluca~de Cola}, \bibinfo{person}{Tristan Deleu}, \bibinfo{person}{Manuel Goulão}, \bibinfo{person}{Andreas Kallinteris}, \bibinfo{person}{Arjun KG}, \bibinfo{person}{Markus Krimmel}, \bibinfo{person}{Rodrigo Perez-Vicente}, \bibinfo{person}{Andrea Pierré}, \bibinfo{person}{Sander Schulhoff}, \bibinfo{person}{Jun~Jet Tai}, \bibinfo{person}{Andrew Tan~Jin Shen}, {and} \bibinfo{person}{Omar~G. Younis}.} \bibinfo{year}{2023}\natexlab{}.
\newblock \bibinfo{title}{Gymnasium}.
\newblock
\newblock
\urldef\tempurl%
\url{https://doi.org/10.5281/zenodo.8127026}
\showDOI{\tempurl}


\bibitem[Wan et~al\mbox{.}(2024)]%
        {wan2024coverage}
\bibfield{author}{\bibinfo{person}{Xiaohui Wan}, \bibinfo{person}{Tiancheng Li}, \bibinfo{person}{Weibin Lin}, \bibinfo{person}{Yi Cai}, {and} \bibinfo{person}{Zheng Zheng}.} \bibinfo{year}{2024}\natexlab{}.
\newblock \showarticletitle{Coverage-guided fuzzing for deep reinforcement learning systems}.
\newblock \bibinfo{journal}{\emph{Journal of Systems and Software}}  \bibinfo{volume}{210} (\bibinfo{year}{2024}), \bibinfo{pages}{111963}.
\newblock


\bibitem[Wardat et~al\mbox{.}(2021)]%
        {wardat2021deeplocalize}
\bibfield{author}{\bibinfo{person}{Mohammad Wardat}, \bibinfo{person}{Wei Le}, {and} \bibinfo{person}{Hridesh Rajan}.} \bibinfo{year}{2021}\natexlab{}.
\newblock \showarticletitle{Deeplocalize: Fault localization for deep neural networks}. In \bibinfo{booktitle}{\emph{2021 IEEE/ACM 43rd International Conference on Software Engineering (ICSE)}}. IEEE, \bibinfo{pages}{251--262}.
\newblock


\bibitem[Watkins and Dayan(1992)]%
        {watkins1992q}
\bibfield{author}{\bibinfo{person}{Christopher~JCH Watkins} {and} \bibinfo{person}{Peter Dayan}.} \bibinfo{year}{1992}\natexlab{}.
\newblock \showarticletitle{Q-learning}.
\newblock \bibinfo{journal}{\emph{Machine learning}}  \bibinfo{volume}{8} (\bibinfo{year}{1992}), \bibinfo{pages}{279--292}.
\newblock


\bibitem[Xu et~al\mbox{.}(2010)]%
        {xu2010method}
\bibfield{author}{\bibinfo{person}{Zhangyan Xu}, \bibinfo{person}{Shichao Shang}, \bibinfo{person}{Wenbin Qian}, {and} \bibinfo{person}{Wenhao Shu}.} \bibinfo{year}{2010}\natexlab{}.
\newblock \showarticletitle{A method for fuzzy risk analysis based on the new similarity of trapezoidal fuzzy numbers}.
\newblock \bibinfo{journal}{\emph{Expert Systems with Applications}} \bibinfo{volume}{37}, \bibinfo{number}{3} (\bibinfo{year}{2010}), \bibinfo{pages}{1920--1927}.
\newblock


\bibitem[Zadeh(1965)]%
        {zadeh1965fuzzy}
\bibfield{author}{\bibinfo{person}{Lotfi~Asker Zadeh}.} \bibinfo{year}{1965}\natexlab{}.
\newblock \showarticletitle{Fuzzy sets}.
\newblock \bibinfo{journal}{\emph{Information and control}} \bibinfo{volume}{8}, \bibinfo{number}{3} (\bibinfo{year}{1965}), \bibinfo{pages}{338--353}.
\newblock


\bibitem[Zhang et~al\mbox{.}(2020)]%
        {zhang2020machine}
\bibfield{author}{\bibinfo{person}{Jie~M Zhang}, \bibinfo{person}{Mark Harman}, \bibinfo{person}{Lei Ma}, {and} \bibinfo{person}{Yang Liu}.} \bibinfo{year}{2020}\natexlab{}.
\newblock \showarticletitle{Machine learning testing: Survey, landscapes and horizons}.
\newblock \bibinfo{journal}{\emph{IEEE Transactions on Software Engineering}} \bibinfo{volume}{48}, \bibinfo{number}{1} (\bibinfo{year}{2020}), \bibinfo{pages}{1--36}.
\newblock


\bibitem[Zheng et~al\mbox{.}(2018)]%
        {zheng2018drn}
\bibfield{author}{\bibinfo{person}{Guanjie Zheng}, \bibinfo{person}{Fuzheng Zhang}, \bibinfo{person}{Zihan Zheng}, \bibinfo{person}{Yang Xiang}, \bibinfo{person}{Nicholas~Jing Yuan}, \bibinfo{person}{Xing Xie}, {and} \bibinfo{person}{Zhenhui Li}.} \bibinfo{year}{2018}\natexlab{}.
\newblock \showarticletitle{DRN: A deep reinforcement learning framework for news recommendation}. In \bibinfo{booktitle}{\emph{Proceedings of the 2018 world wide web conference}}. \bibinfo{pages}{167--176}.
\newblock


\bibitem[Zolfagharian et~al\mbox{.}(2023)]%
        {zolfagharian2023search}
\bibfield{author}{\bibinfo{person}{Amirhossein Zolfagharian}, \bibinfo{person}{Manel Abdellatif}, \bibinfo{person}{Lionel~C Briand}, \bibinfo{person}{Mojtaba Bagherzadeh}, {and} \bibinfo{person}{S Ramesh}.} \bibinfo{year}{2023}\natexlab{}.
\newblock \showarticletitle{A search-based testing approach for deep reinforcement learning agents}.
\newblock \bibinfo{journal}{\emph{IEEE Transactions on Software Engineering}} (\bibinfo{year}{2023}).
\newblock


\end{thebibliography}

\end{document}